\documentclass[reprint,superscriptaddress,prb,floatfix]{revtex4-2}

\usepackage[T1]{fontenc}
\usepackage{graphicx}% Include figure files
\usepackage{dcolumn}% Align table columns on decimal point
\usepackage{siunitx}
\DeclareSIUnit{\barpressure}{bar}
\DeclareSIUnit{\langmuir}{L}
\DeclareSIUnit\angstrom{\protect \text {Å}}
\DeclareSIUnit\rydberg{\text{Ry}}
\usepackage{amsmath}
\usepackage{xcolor}
\usepackage{comment}
\usepackage{physics}
\usepackage{ulem}
\newcommand\quotemarks[1]{‘‘#1’’}
\newcommand\hl[1]{%
#1}%
\begin{document}
%TC:ignore
\preprint{APS/123-QED}

\title{Orbital texture of valence band and interband transitions of the type-II multiferroic van der Waals NiI\textsubscript{2}}

\author{A. De Vita$^\$$}
\email{alessandro.de.vita@tu-berlin.de}
\affiliation{Institut für Physik und Astronomie, Technische Universität Berlin, Straße des 17 Juni 135, 10623 Berlin, Germany\looseness=-1}%
\affiliation{Fritz Haber Institute of the Max Planck Society, Faradayweg 4--6, 14195 Berlin, Germany\looseness=-1}%
\author{O.E. Giehl$^\$$}
\affiliation{Institut für Physik und Astronomie, Technische Universität Berlin, Straße des 17 Juni 135, 10623 Berlin, Germany\looseness=-1}%
\affiliation{Fritz Haber Institute of the Max Planck Society, Faradayweg 4--6, 14195 Berlin, Germany\looseness=-1}%
\author{T. Mukherjee}
\affiliation{Institut für Physik und Astronomie, Technische Universität Berlin, Straße des 17 Juni 135, 10623 Berlin, Germany\looseness=-1}%
\affiliation{Fritz Haber Institute of the Max Planck Society, Faradayweg 4--6, 14195 Berlin, Germany\looseness=-1}%
\author{M. Wolf}
\affiliation{Fritz Haber Institute of the Max Planck Society, Faradayweg 4--6, 14195 Berlin, Germany\looseness=-1}%
\author{L. Rettig}
\affiliation{Fritz Haber Institute of the Max Planck Society, Faradayweg 4--6, 14195 Berlin, Germany\looseness=-1}%
\affiliation{Department of Physics, RPTU University Kaiserslautern-Landau, 67663 Kaiserslautern, Germany\looseness=-1}%
\affiliation{OPTIMAS Research Center, RPTU University Kaiserslautern-Landau, 67663 Kaiserslautern, Germany\looseness=-1}%
\author{R. Ernstorfer}
\affiliation{Institut für Physik und Astronomie, Technische Universität Berlin, Straße des 17 Juni 135, 10623 Berlin, Germany\looseness=-1}%
\affiliation{Fritz Haber Institute of the Max Planck Society, Faradayweg 4--6, 14195 Berlin, Germany\looseness=-1}%
\author{T. Pincelli}
\email{tommaso.pincelli@tu-berlin.de}
\affiliation{Institut für Physik und Astronomie, Technische Universität Berlin, Straße des 17 Juni 135, 10623 Berlin, Germany\looseness=-1}%
\affiliation{Fritz Haber Institute of the Max Planck Society, Faradayweg 4--6, 14195 Berlin, Germany\looseness=-1}%
\author{L. T. Lloyd}
\affiliation{Fritz Haber Institute of the Max Planck Society, Faradayweg 4--6, 14195 Berlin, Germany\looseness=-1}%

%\keywords{Van der Waals, quantum materials, ARPES, magnetism, CrI3}

\date{\today}% It is always \today, today,
%\preprint{APS/123-QED}

\begin{abstract}

NiI$_2$ is a van der Waals semiconductor characterized by complex electronic and magnetic phases, holding promise towards functional applications. For this purpose, a detailed understanding of its band structure and its excitations is essential to elucidate the electronic and transport properties. We present a comprehensive description of the band structure and of the lowest valence-to-conduction band transition in bulk NiI$_2$, by means of time- and angle-resolved photoemission spectroscopy supported by first-principles calculations. We describe the orbital character of Ni and I manifolds, and identify the main momentum-resolved features of the valence band and the shallow minima of the conduction band. With this study, we provide a foundational understanding of the electronic structure of NiI$_2$ and its single-particle interband electronic excitation, a benchmark for theoretical models, and a framework for future investigations on this material.

\end{abstract}

%TC:endignore

\maketitle

\noindent$^\$$A.D.V. and O.E.G. contributed equally to this work.

\section{Introduction}

Van der Waals (vdW) materials have attracted considerable attention due to a plethora of emergent properties at low dimensionality such as increased correlations, orbital reordering, and unconventional magnetism \cite{Samarth2017,Gibertini2019,Kajale2024}. In particular, the quest for functional materials in the thriving field of spintronics demands electrically-controllable magnetic configurations. The transition-metal halide NiI$_2$ \cite{McGuire2017} appears as an ideal candidate thanks to the strong coupling between electronic and magnetic degrees of freedom: below $\sim\SI{59}{\K}$ the crystal enters a helimagnetic antiferromagnetic phase and a ferroelectric polarization perpendicular to the helix propagation vector appears. NiI$_2$ is thus a type-II multiferroic vdW material, and displays a robust magnetoelectric response down to the monolayer (ML) limit \cite{Kurumaji2013,Song2022b,Song2025,Pan2026,Cuono2026}. It has also been proposed that NiI$_2$ hosts an unconventional high-spin Kitaev state, at the origin of quantum-spin-liquid states and topological spin textures \cite{Wang2026,Kim2026}.

It is therefore surprising that only one spectroscopic study aimed at describing its low-energy electronic properties, in terms of valence and conduction bands, has been previously performed \cite{Starnberg1986}, while photoelectron spectroscopies are ideal to fill this gap. Important technical challenges are related to the chemical stability and conductivity of NiI$_2$, which, similarly to other compounds in the halide family \cite{DeVita2022,Sant2023,DeVita2026,DeVita2026b}, is extremely hygroscopic and displays significant charging upon photoemission, requiring ultrahigh vacuum integration and charging compensation strategies. 

In this work, we describe in detail the valence band properties and the lowest momentum-resolved interband transitions of NiI$_2$, combining time- and angle-resolved photoemission spectroscopy (trARPES) with density functional theory (DFT) calculations. By comparing experiments with theory, we identify the orbital character of electronic states. Upon optical excitation we observe the appearance of interband transitions, whose momentum distribution reveals the texture of the shallow conduction band (CB). Our findings provide essential information on the electronic structure of this material, and pave the way for further spectroscopic investigations in view of a functionalization of NiI$_2$ in the field of opto-spintronics.

\section{Methods}

The experiment has been performed using the high-harmonics generation (HHG) setup in the ARPES laboratory at the Fritz Haber Institute of the Max Planck Society, Berlin \cite{Puppin2019,Maklar2020}. Static ARPES spectra have been acquired with XUV photons with variable polarization at \SI{21.7}{\eV}; for polarization-integrated measurements, the probe polarization was controlled via a half-wave plate, which spanned the whole range from \ang{0} to \ang{360}, and the collected ARPES intensity was summed over all half-wave plate angles. Time-resolved data, integrated from pump-probe overlap to \SI{1}{\pico\s}, have been measured by pumping the sample with \SI{1.91}{\eV} \textit{p}-polarized pulses from an optical parametric amplifier (OPA), and using XUV \textit{p}-polarized photons at \SI{21.7}{\eV} as probe; the pump fluence was \SI{60}{\micro\J\per\cm^2}. The sample has been measured at \SI{300}{\K}. ARPES spectra have been collected with a TOF momentum microscope at \ang{65} incidence angle and normal emission conditions; measurements have been perfomed in a regime with $<1$ detected electron per pulse, to minimize space charge effects. The overall energy resolution has been estimated to be $\approx\SI{100}{\milli\eV}$; the FWHM of the pump (probe) pulse is $\approx\SI{50}{\fs}$ ($\approx\SI{100}{\fs}$). \hl{Pump-probe spatial overlap has been determined in PEEM mode by aligning the pump spot in front of the aperture of the momentum microscope.} The pump (probe) spot size is \qtyproduct{150x150}{\micro\metre} (\qtyproduct{80x80}{\micro\metre}) FWHM. Commercially available NiI\textsubscript{2} crystals (HQ Graphene B.V.) have been mounted on sample holders using silver epoxy glue (HD20E, Epotek), and cleaved in UHV ($p<\SI{8e-11}{\milli\barpressure}$) before the measurement. To compensate for sample charging, a \SI{400}{\nano\metre}, \hl{\SI{400}{\milli\W}} continuous wave (CW) laser has been used in conjunction with XUV during the measurement. Over the course of a \hl{single acquisition ($\sim$one day of measurement)}, we notice no evolution of the lineshape of the valence band (Supporting Information, Fig.~S1). \hl{To check the absence of a photoemission signal induced by the CW laser, we compared EDCs at negative and positive delays, as described in the Supporting Information, Fig.~S2.}

The energy axis in all the (\textit{E}, \textit{k}) spectra has been aligned such that the valence band maximum (VBM) corresponds to zero energy. We employed a standard approach in semiconductor physics, as described in Refs. \cite{Kraut1980,Katnani1983}: we linearly extrapolate the spectral leading edge to the baseline, and take the intersection as a proxy for VBM. \hl{The systematic error in the determination of the VBM via the fitting procedure is calculated to be \SI{2}{\milli\eV}.} 

DFT calculations were performed using the Quantum ESPRESSO suite~\cite{Giannozzi_2009, Giannozzi_2017, Giannozzi_2020} with projector augmented wave (PAW) pseudopotentials~\cite{PAW} and the Perdew-Burke-Ernzerhof (PBE) parametrization of the generalized gradient approximation (GGA) exchange-correlation functional~\cite{PBE1, PBE2} \hl{with ortho-atomic projectors}. A plane-wave cutoff energy of $\SI{90}{\rydberg}$ was used. In order to account for correlation effects in the Ni 3\textit{d} orbitals, a DFT+\textit{U}+\textit{J} approach as introduced by Liechtenstein \textit{et al.}~\cite{Liechtenstein} was employed, and several combinations of different \textit{U} and \textit{J} values were tested to find the best match with the experiment.

NiI$_2$ was simulated using a hexagonal unit cell with lattice constants of $a=\SI{3.92}{\angstrom}$ and $c=\SI{19.80}{\angstrom}$ and a Ni-I bond length of $\SI{2.8}{\angstrom}$~\cite{LatticeParams-French, LatticeParams-SpringerMat}\hl{, without structure relaxation}. \hl{The magnetic order of NiI$_2$ is modelled using a ferromagnetic (FM) ordering for each layer of NiI$_2$ with an antiferromagnetic (AFM) inter-layer order~\cite{EffectOfPressure, VaporDeposition, PhotocurrentSpectroscopy}}. 
For ML calculations, a vacuum layer $>\SI{19}{\angstrom}$ was used between MLs to prevent unwanted inter-layer coupling. For bulk calculations, a $1\times1\times2$ supercell was used to accommodate the AFM order of bulk NiI$_2$. An $8\times8\times1$ $k$-mesh centered on $\Gamma$ was used for the BZ sampling in ML calculations, and an $8\times8\times3$ $k$-mesh for bulk calculations. \hl{We ran test calculations up to a $12\times12\times3$ mesh in the bulk case and $18\times18\times1$ in the ML case: the dispersion and orbital character of the bands did not change when increasing the number of \textit{k}-points. We decided for $8\times8$ meshes as a practical choice: The lower number of k-points results in faster computation times while still being accurate to our own tests at denser \textit{k}-meshes and to published literature. Momentum-resolved maps were generated by adding a \SI{0.065}{\rydberg} energy broadening.}

\hl{Due to the surface sensitivity of ARPES measurements, we employ the surface notation for the high-symmetry points ($\bar{\mathrm{K}}-\bar{\mathrm{\Gamma}}-\bar{\mathrm{M}}$); DFT ML calculations refer to high-symmetry points without the bar notation ($\mathrm{K}-\Gamma-\mathrm{M}$).}

\begin{figure*}[htb]
\includegraphics[width=0.95\linewidth]{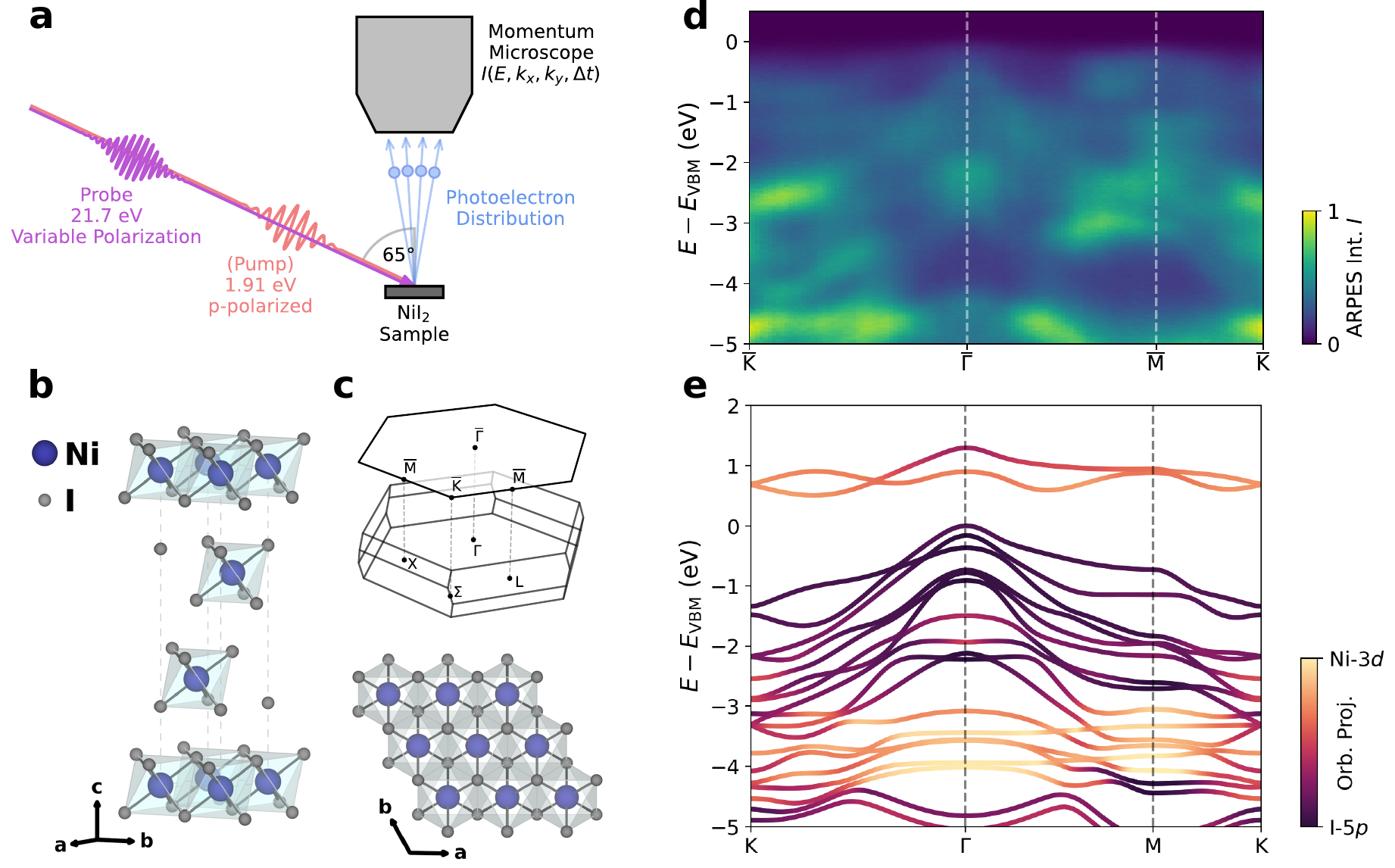}
\caption{\protect\label{fig:Fig1} \textbf{a}. Cartoon of the experimental setup and geometry for tr-ARPES measurements. A \textit{p}-polarized pump at \SI{1.91}{\eV} impinges on the sample at \ang{65} and is followed by a collinear probe pulse in the XUV range (\SI{21.7}{\eV}) with variable polarization; photoelectrons are collected into a TOF-momentum microscope where their (\textit{E}, \textit{k}) distribution is measured. \textbf{b}. Top and side hard sphere view of NiI\textsubscript{2} crystal structure. \textbf{c}. Bulk and surface Brillouin zone of NiI$_2$, with explicit labeling of the main high-symmetry points. \textbf{d}. ARPES spectrum of  NiI\textsubscript{2} valence band dispersion along high-symmetry lines, measured at $T=\SI{300}{K}$ and integrated over light polarization. \textbf{e}. Band structure dispersion along the same high-symmetry lines, as calculated by DFT. The color scale identifies the orbital contribution of the atomic species to the band structure. Panel \textbf{d} and \textbf{e} are aligned such that, even though the total energy range is different \hl{to accommodate for the inclusion of the CB}, energy intervals are equally spaced in the image.}
\end{figure*}

\section{Results and discussion}

NiI$_2$ crystallizes in the rhombohedral ($\mathrm{R\bar{3}m}$) crystal structure (space group 166), as depicted together with the bulk and surface Brillouin zone (BZ) in Fig.~\ref{fig:Fig1}b-c. The magnetic transition at $\sim\SI{59}{\K}$ drives the system into a helimagnetic antiferromagnetic state, where (as a result of strong spin-orbit coupling) inversion symmetry is broken, but inversion symmetry is preserved in the bulk in the nonmagnetic state. The ARPES (\textit{E}, \textit{k}) spectrum of NiI$_2$ at \SI{300}{\K} along the $\bar{\mathrm{K}}-\bar{\mathrm{\Gamma}}-\bar{\mathrm{M}}-\bar{\mathrm{K}}$ high-symmetry directions of the surface BZ are shown in Fig.~\ref{fig:Fig1}d. Despite the relatively high temperature of the ARPES measurements, band structure features and dispersions are easily identifiable, confirming the high crystalline quality of the sample surface. To understand the orbital character of these bands, we have modeled the electronic structure using non-collinear DFT including spin-orbit coupling (SOC). Due to the vdW nature of bulk NiI$_2$ and similarly to other vdW magnets \cite{Nitschke2025,Nitschke2026}, we expect the band structure to be dominated by intralayer interactions, with interlayer interactions playing a minor role; therefore we model NiI$_2$ as a ML \cite{Lebedev2024, Song2025, EffectOfPressure, VaporDeposition}. The resulting calculated band structure is shown in Fig.~\ref{fig:Fig1}e. Based off these results, we can identify the valence band maximum (VBM) to be dominated by strongly dispersive iodine 5\textit{p} states, extending for several electronvolts of bandwidth. We also reveal weakly dispersive states from \SIrange{-1.5}{-1.9}{\eV}, originating from hybridized I 5\textit{p} -- Ni 3\textit{d} orbitals; similarly, weakly dispersive states with mostly Ni 3\textit{d} character are localized in the range from \SIrange{-3.5}{-5}{\eV}. The lowest CB also appears to be underpinned by Ni orbitals, supporting the charge transfer insulating nature of the bandgap \cite{Occhialini2024}.

We systematically evaluated various Hubbard \textit{U} and Hund's coupling \textit{J} parameters for the Ni 3\textit{d} manifolds, as described in Fig.~S3 of the Supporting Information \cite{SupplementalMaterial}. The best match was found with $U=\SI{3.5}{\eV}$ and $J=\SI{0.6}{\eV}$. The nominal $d^8$ configuration of the Ni ions means that two unpaired electrons are found in a high-spin configuration ($S=1$); to minimize the total energy, these electrons moderately delocalize (likely helped by metal-ligand hybridization) and stabilize a Hund's energy gain. Hence, a non-zero \textit{J} is reasonable. A Hubbard $U=\SI{3.5}{\eV}$ is in line with a charge-transfer-type bandgap, such that $\Delta<U$. The chosen combination of parameters yields a DFT \hl{direct} bandgap of $\sim$\SI{0.9}{\eV}. This value is at odds with the optical bandgap (interband absorption edge), experimentally determined to be \SI{1.6}{\eV} \cite{Lebedev2024}; the underestimation of the bandgap is a known shortcoming of DFT calculations. Overall, ML DFT calculations capture the main features of the band dispersion in NiI$_2$, in particular the highly dispersive nature of I 5\textit{p} states along $\Gamma$-K and $\Gamma$-M, the deeper regions of the valence band (from \SIrange{-4}{-5}{\eV}), and the energy position and dispersion of the higher intensity hotspots near the K and M symmetry points, at \SI{-2.7}{\eV} and \SI{-3}{\eV} respectively. We note that, while the energy-momentum dispersion can be compared between ARPES measurements and DFT calculations, the intensity of the ARPES signal arises from many-body interactions and Bloch wavefunction symmetries affecting photoemission matrix elements, which are not accounted for in the calculations. For example, our data do not reveal some of the bands localized at $\sim\SI{-4}{\eV}$ at $\Gamma$, probably because of unfavorable matrix elements suppressing the photoemitted intensity.

As already mentioned, NiI$_2$ is a vdW material, therefore it is expected that interlayer electronic couplings are weak. To quantitatively evaluate which contributions to the band structure might originate from layer-layer interactions, we performed collinear DFT calculations modelling bulk and ML NiI$_2$ (Supporting Information, Fig.~S4). A comparison between the two calculations reveals only minor differences in the electronic structure. The only sizable discrepancy is found when observing the bundle of bands between \SI{-2}{\eV} and \SI{-3}{\eV} in the ARPES spectra: bulk calculations display a higher spectral density located in this energy-momentum region. \hl{To further inquire into the orbital nature of the bands, we projected the DFT bands onto the atomic orbitals for both ML and bulk NiI$_2$, as displayed in Supporting Information, Fig.~S5. These calculations show that the VBM displays predominantly $p_x/p_y$ character, with a weak $p_z$ component emerging from \SI{-0.5}{\eV}. Conversely, at the center of the BZ and at higher binding energies, bands in the ML calculations are mostly originating from $p_z$ orbitals for I, and $d$ orbitals with strong $z$ component for Ni. This fact is even more evident in bulk calculations, especially in the region between \SI{-2}{\eV} and \SI{-3}{\eV} where ML calculations display no intensity at $\Gamma$, whereas in the bulk most of the bands are $p_z$ or $d_{zy}$ character. While the BZ center is somewhat depleted of $p_x/p_y$ contributions, close to the K and M points there is a higher density of bands with in-plane character ($p_x$, $p_y$, $d_{x^2-y^2}$). Lastly, the CB is composed of mixed in-plane/out-of-plane character for both Ni and I species. These considerations are in good accord with} polarization-resolved ARPES measurements (Supporting Information, Fig.~S6) \hl{. Indeed, the VBM is not clearly visible with \textit{p}-polarized light, but becomes spectrally evident with \textit{s}-polarized light; bands between \SI{-2}{\eV} and \SI{-3}{\eV} are revealed in \textit{p}-polarized spectra, similarly to bands between \SI{-4}{\eV} and \SI{-5}{\eV} where a large share of the $d_{z^2}$ contribution is located; and a higher ARPES intensity is measured outside of the BZ center in the \textit{s}-polarized scan.}

\begin{figure*}[htb]
\includegraphics[width=0.95\linewidth]{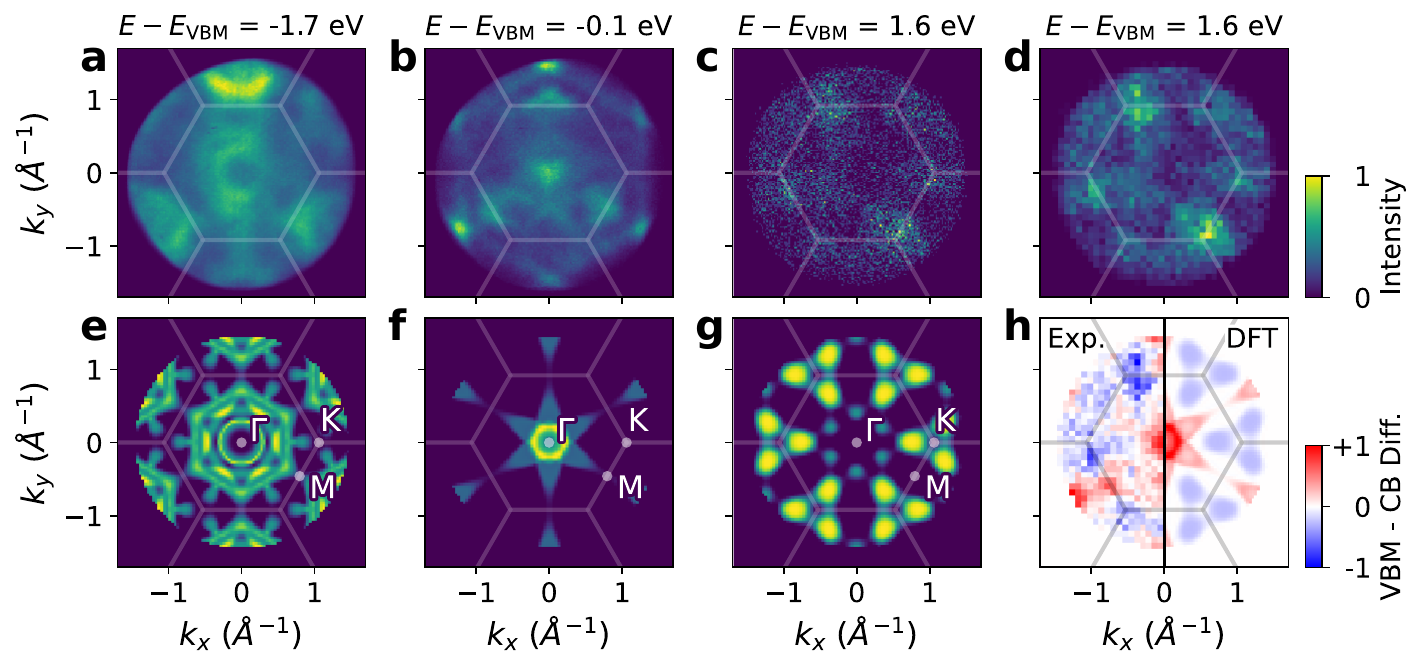}
\centering
\caption{\protect\label{fig:Fig2}\textbf{a}, \textbf{b}. Isoenergetic ARPES valence band momentum maps, at energies $E-E_\text{VBM}=\SI{-1.7}{\eV}$ and $E-E_\text{VBM}=\SI{-0.1}{\eV}$ \hl{integrated over a \SI{0.1}{\eV} window, at $T=\SI{300}{K}$}. The white overlay highlights the contours of the two-dimensional surface Brillouin zone. \textbf{c}. Raw isoenergetic ARPES momentum maps of the excited state, at the energy \SI{1.6}{\eV}\hl{, integrated over a \SI{0.3}{\eV} window and at $T=\SI{300}{K}$}. \textbf{d}. Same as \textbf{c}, but data have been binned with a factor 5 to facilitate visualization. \textbf{e}, \textbf{f}. DFT calculations of the isoenergetic momentum maps corresponding to the valence band features in \textbf{a} and \textbf{b}. \textbf{g}. DFT calculation of the isoenergetic momentum map of the bottom of the CB. \textbf{h}. \textit{Left}: Difference plot between the isoenergetic ARPES momentum map at the VBM (Fig.~\ref{fig:Fig2}b) and at the excited state (Fig.~\ref{fig:Fig2}c). \textit{Right}: Difference plot between the calculated isoenergetic momentum map at the VBM (Fig.~\ref{fig:Fig2}f) and at the CBM (Fig.~\ref{fig:Fig2}g). The color scale in \textbf{a}, \textbf{b}, \textbf{c}, and \textbf{d} corresponds to ARPES intensity; in \textbf{e}, \textbf{f}, and \textbf{g} it represents the DOS normalized intensity.}
\end{figure*}

Having identified the main orbital-resolved features of the valence band structure, we move to the characterization of the momentum-resolved character. Fig.~\ref{fig:Fig2}a-b shows the isoenergetic ($k_x$, $k_y$) momentum maps \hl{at the shallow I-Ni hybridized manifold and at the VBM, respectively}. The ARPES data display a petal-like texture at the VBM, with pockets around $\bar{\Gamma}$ and close to three $\bar{\mathrm{M}}$ points; adjacent $\bar{\mathrm{M}}$ and $\bar{\mathrm{K}}$ points are spectrally inequivalent, differently from what is expected from the hexagonal surface BZ. At \SI{-1.7}{\eV} binding energy the threefold symmetry is even more evident in the three pairs of spokes departing from the concentric circles centered at $\bar{\Gamma}$. The experimental momentum maps can be compared with DFT calculations at the corresponding energies, displayed in Fig.~\ref{fig:Fig2}e-f. Theoretical results replicate the main features of the momentum distribution of the valence band, including the pockets around $\Gamma$ at \SI{-1.7}{\eV} (Fig.~\ref{fig:Fig2}a,e) and the spikes close to the BZ edges at \SI{-0.1}{\eV} (Fig.~\ref{fig:Fig2}b,f); but one notable feature is that DFT calculations show a sixfold symmetry, differently from ARPES maps.

In NiI$_2$, a potential source of inversion-symmetry breaking is connected to the spin-spiral order \cite{Cuono2026}, such that the electronic states at $+\mathbf{k}$ and $-\mathbf{k}$ are inequivalent in the magnetic state. \hl{Measurements in Figs.~\ref{fig:Fig1} and \ref{fig:Fig2} have been performed above the magnetic ordering temperature, and inversion symmetry in the bulk should be preserved. To check the effect of magnetic order upon the electronic band structure, we have performed a static ARPES measurement at \SI{30}{\K}, below the magnetic ordering temperature (Fig.~S7). By comparing the results at different temperature, we can note that, apart from an increased sharpness in the electronic features due to the lower temperature, no macroscopic changes in terms of gap openings or band renormalizations appear, and the symmetry of the momentum maps appears unaffected. Moreover, we detect no stabilization of an in-plane supercell by ARPES within our resolution, which is consistent with the antiferromagnetic order lying mostly along the \textit{c} axis ($q=[0.087,0.087,1.5]$ \cite{Tseng2025}). The similarities between the band structures above and below the magnetic transition supports the comparison of ARPES spectra with antiferromagnetic DFT+\textit{U}+\textit{J}, consistently with examples in the family of vdW antiferromagnets \cite{Sternemann2026,Nitschke2026}. Furthermore, the absence of modifications to the whole band structure suggests that localized Ni moments dominate the magnetic interactions, such that contributions from interatomic hybridization are minor; therefore, we do not expect magnetic correlations persisting above the magnetic transitions to be at the origin of the reduced symmetry.} In our case, the most likely explanation is broken inversion symmetry at the surface. Similarly to inversion-symmetric transition metal dichalcogenides, inversion symmetry is present in the bulk space group but not in every site point group \cite{Zhang2014}. Individual subsets of the BZ are thus inversion-asymmetric: ARPES preferentially probes the topmost layer of the unit cell, and thus we observe a threefold symmetry of photoemission intensity due to the hidden symmetry breaking of individual sites, revealed at the surface. In general, the variation of matrix elements generated by the experimental asymmetry (Fig.~\ref{fig:Fig1}a) can also introduce a modulation of the photoemitted intensity \cite{Daimon1993,Boban2025}. Integrating over all XUV polarization directions, as we did in our measurement, mitigates this effect, and thus we would expect it to be a minor correction, and not the main cause for the significant intensity variation we observe.

By optically pumping the sample with \SI{1.91}{\eV} photons, \textit{i.e.}, above the optical bandgap, we were able to reveal a photoemission signal from the excited states and, thanks to the access to the full momentum-resolved distribution granted by trARPES, its texture in reciprocal space. The momentum map shown in Fig.~\ref{fig:Fig2}c displays the \textit{k}-dependent distribution of ARPES intensity at \SI[uncertainty-mode = separate]{1.6\pm0.3}{\eV} above the VBM, probed after excitation. The measurement exhibits a momentum distribution with high photoemitted intensity at the outskirts of the BZ, mostly localized near the K points, whereas the signal at the center of the ($k_x$, $k_y$) distribution of the excited signal is mostly absent. This is made more evident to the eye if the raw data in Fig.~\ref{fig:Fig2}c are binned to enhance the signal to noise ratio, as done in Fig.~\ref{fig:Fig2}d.

\begin{figure*}[htb]
\includegraphics[width=0.95\linewidth]{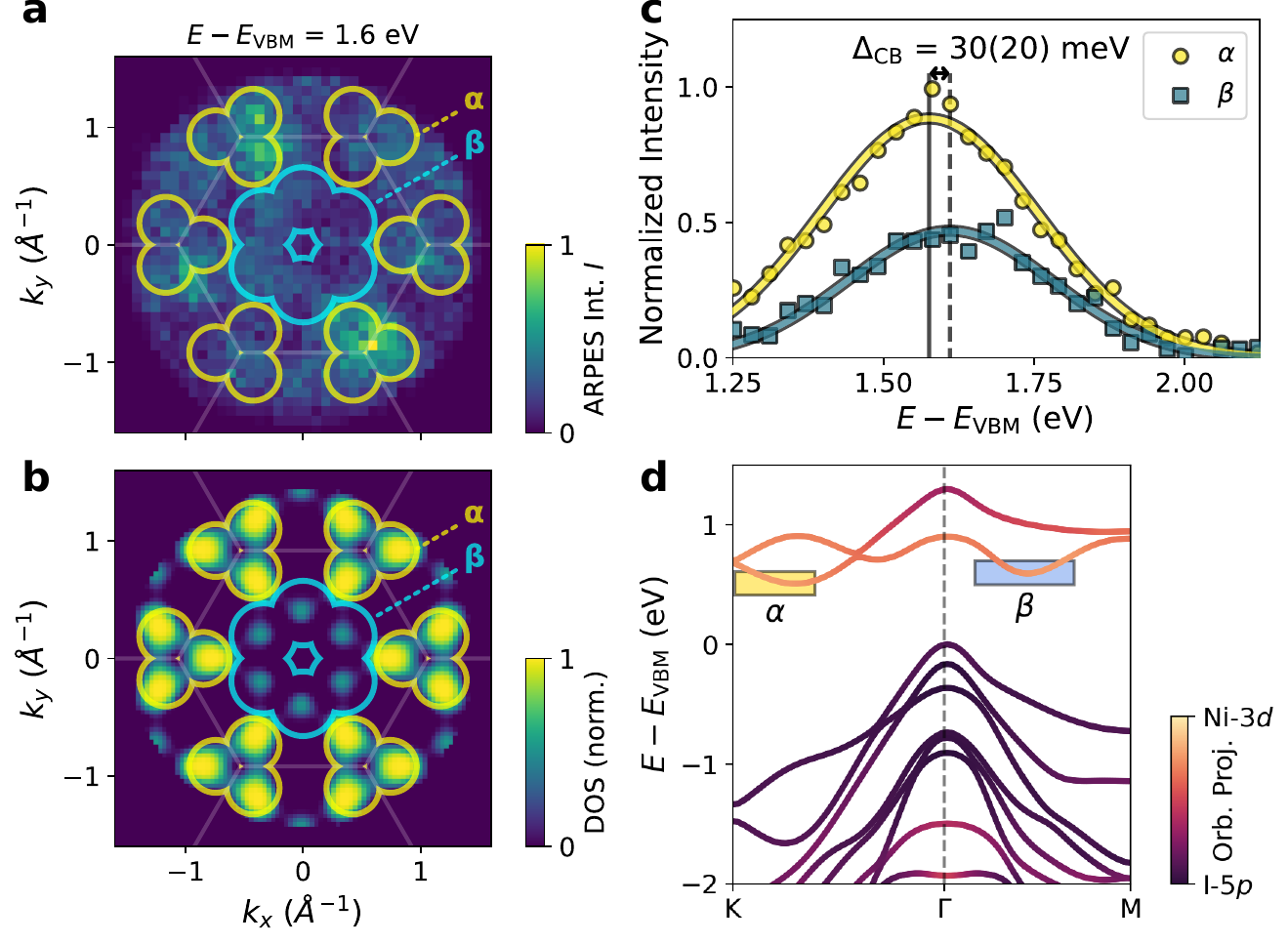}
\centering
\caption{\protect\label{fig:Fig3}\textbf{a}. Isoenergetic ARPES momentum map of the excited state \hl{at $T=\SI{300}{K}$}, with overlayed masking of two momentum-space regions labeled $\alpha$ and $\beta$. \textbf{b}. Corresponding DFT calculation of the isoenergetic momentum map of the CBM, with the same masking overlayed. \textbf{c}. Energy distribution curves of regions $\alpha$ and $\beta$ in the energy range of the excited state; solid lines represent the fitting of the experimental points, from which the energy difference between the peaks has been extracted. \textbf{d}. DFT calculations showing the VBM-adjacent region together with the conduction band. The squares indicate the band structure areas mapped by the regions $\alpha$ and $\beta$.}
\end{figure*}

There are strong similarities between the experimental features in the momentum map in Fig.~\ref{fig:Fig2}c and the momentum distribution of the density of states (DOS) intensity as calculated by DFT in correspondence of the shallow CB minima (Fig.~\ref{fig:Fig2}g): intense pockets are located near the K points, and secondary features appear along the $\Gamma$-M direction. Moreover, the energy position of the ARPES intensity agrees with the expected optical bandgap for NiI$_2$. These factors hint at an identification of this signal with the CB, where electrons are directly excited by the above-bandgap pump pulse. To confirm this interpretation, we display the difference between the momentum distribution of the VBM (Fig.~\ref{fig:Fig2}a) and that of the excited state (Fig.~\ref{fig:Fig2}c) in the left half of Fig.~\ref{fig:Fig2}h. We note that the signal from the excited state dominates the intensity near the K points, whereas the VBM is emphasized at the center of the BZ and along the $\bar{\Gamma}$-$\bar{\mathrm{M}}$ directions. The difference plot is in excellent agreement with corresponding calculations of the difference between the DOS intensity calculated at the VBM and at the excited state, plotted in the right half of Fig.~\ref{fig:Fig2}h. \hl{Directly comparing the excited-state signal with the DFT-calculated DOS by means of momentum-distribution curves (MDCs) also shows a good match between experiment and theory (Supporting Information, Fig.~S8). To check whether the excited states are dependent upon the pump polarization, we repeated the measurement with \textit{s}-polarized pump; results are reported in Fig.~S9 of the Supporting Information, where it can be seen how both pump polarizations display the same momentum distribution and the same relaxation after time overlap. The similarity between the two conditions excludes any spurious influence of laser-assisted photoemission effects.} Therefore, we may conclude that the excited state signal is a signature of single-particle states arising from interband transition from the VBM to the CB minimum (CBM).

The VBM is composed mainly of I 5\textit{p} orbitals, whereas the CB arises from Ni 3\textit{d} states. By identifying the CB minima via their reciprocal space signatures, we confirm the charge-transfer nature of the bandgap, whereupon the interband transition involves I \textit{p} to Ni \textit{d} excitations.

As mentioned when discussing DFT calculations in Fig.~\ref{fig:Fig2}g, the momentum texture of the CB minima comprises two features, surrounding the K points and along the $\Gamma$-M direction. Fig.~\ref{fig:Fig3} elucidates the role of different momentum areas in revealing the texture of the excited state. We mask distinct regions of the momentum space as shown in Fig.~\ref{fig:Fig3}a, following the expected distribution of the excited state based off the calculated momentum map (shown in Fig.~\ref{fig:Fig3}b with the same masking). We then extract the corresponding energy distribution curves (EDC), where the signal of the excited state is located (Fig.~\ref{fig:Fig3}c), and we fit the peaks of photoemitted intensity \hl{with Gaussian fits}. The fit reveals an energy separation $\Delta_{CB}=\SI{30\pm20}{\milli\eV}$ between the peaks. Such a difference, small but noticeable, reflects the distinct features in the CB manifold, as illustrated in Fig.~\ref{fig:Fig3}d. The regions labeled \quotemarks{$\alpha$}, corresponding to interband transitions towards the CBM, are found at lower energies; the area labeled \quotemarks{$\beta$} is a local minimum along $\Gamma$-M and lies at higher energies. The excited state signal, as revealed by the momentum-resolved maps, is therefore mostly composed of two main contributions: our experiment reveals the CBM in region \quotemarks{$\alpha$} at \SI{1.58\pm0.01}{\eV}, and another interband transition towards a different \textit{k}-space region in the CB (\quotemarks{$\beta$}) at \SI{1.61\pm0.02}{\eV}. This result \hl{shows good agreement with the measured interband absorption edge \cite{Lebedev2024}. The extracted $\Delta_{CB}$ underestimates the DFT predictions in terms of the relative energy positions of the CB electron pockets ($\Delta_{DFT}=\SI{86}{\milli\eV}$), likely due to the temperature-induced broadening making the electronic states less energy- and momentum-coherent; however, the momentum texture of the calculated CB pockets qualitatively captures the experimental features revealed by ARPES}. 

In this work, we focus on the energetics and the orbital character of the valence band and the shallow minima of the CB. We do not discuss the dynamics of the excited states to avoid diluting the main message of this work; such a study requires a more in-depth investigation and will be reported elsewhere.

\section{Conclusions}

In summary, we presented a combined experimental and theoretical study of the valence band and interband transitions of NiI$_2$ at room temperature. By comparing static ARPES spectra with DFT calculations, we were able to assess the orbital-resolved band ordering. Time-resolved ARPES allowed us to measure the energy and momentum texture of the excited state, revealing the \textit{k}-space distribution of the interband transitions and identifying the signature of two shallow CB minima. 

Our results provide insight into the electronic structure of NiI$_2$ and enhance the understanding of the orbital character and the \hl{across-bandgap} transitions in this material. By directly visualizing the electronic states, we support the picture of a charge-transfer correlated semiconductor and establish the momentum features of the excited states upon optical pumping. Building upon these results, we expect that future experimental endeavors will \hl{explain thoroughly} the effect of magnetic order upon the electronic states. Moreover, the characterization of the \hl{momentum} texture of the \hl{excited states} \hl{involved in the charge-transfer-gap} interband excitation will serve as a basis for the understanding of optical transitions, and the possible emergence  of bound quasiparticles (\textit{e.g.}, excitons) in NiI$_2$, paving the way towards a full description of its fundamental excitations.

%TC:ignore

\section*{Acknowledgments}
This work is financially supported by the Deutsche Forschungsgemeinschaft (DFG) within Transregio TRR 227 Ultrafast Spin Dynamics project B07, the Max Planck Society, and BERLIN QUANTUM, an initiative endowed by the Innovation Promotion Fund of the city of Berlin.  L.T.L. acknowledges financial support from the Alexander von Humboldt Foundation.
The authors gratefully thank N. Tombros for support in the acquisition and handling of NiI$_2$ samples, and M. Cinchetti and L. Sternemann for valuable insight and fruitful discussion.

\section*{Author Contributions}

L.T.L. and A.D.V conceived the research project. A.D.V., L.T.L., T.M., and O.E.G. performed the experiments. L.T.L., A.D.V. and O.E.G. analyzed the experimental data. DFT calculations were conducted by O.E.G. under the supervision of A.D.V. and T.M. L.R. participated in maintaining and running the experimental apparatus. M.W., R.E., and T.P. provided funding and scientific supervision for the experiments. A.D.V. and O.E.G. wrote the manuscript, with contributions from all co-authors. 

\section*{Data availability}

The data that support the findings of this article are available from the authors upon reasonable request.

\bibliography{bibliography}
%TC:endignore

\end{document}

%\preprint{APS/123-QED}

\title{Supporting Information: Orbital texture of valence band and interband transitions of the type-II multiferroic van der Waals NiI\textsubscript{2}}

\author{A. De Vita$^\$$}
\email{alessandro.de.vita@tu-berlin.de}
\affiliation{Institut für Physik und Astronomie, Technische Universität Berlin, Straße des 17 Juni 135, 10623 Berlin, Germany\looseness=-1}%
\affiliation{Fritz Haber Institute of the Max Planck Society, Faradayweg 4--6, 14195 Berlin, Germany\looseness=-1}%
\author{O.E. Giehl$^\$$}
\affiliation{Institut für Physik und Astronomie, Technische Universität Berlin, Straße des 17 Juni 135, 10623 Berlin, Germany\looseness=-1}%
\affiliation{Fritz Haber Institute of the Max Planck Society, Faradayweg 4--6, 14195 Berlin, Germany\looseness=-1}%
\author{T. Mukherjee}
\affiliation{Institut für Physik und Astronomie, Technische Universität Berlin, Straße des 17 Juni 135, 10623 Berlin, Germany\looseness=-1}%
\affiliation{Fritz Haber Institute of the Max Planck Society, Faradayweg 4--6, 14195 Berlin, Germany\looseness=-1}%
\author{M. Wolf}
\affiliation{Fritz Haber Institute of the Max Planck Society, Faradayweg 4--6, 14195 Berlin, Germany\looseness=-1}%
\author{L. Rettig}
\affiliation{Fritz Haber Institute of the Max Planck Society, Faradayweg 4--6, 14195 Berlin, Germany\looseness=-1}%
\affiliation{Department of Physics, RPTU University Kaiserslautern-Landau, 67663 Kaiserslautern, Germany\looseness=-1}%
\affiliation{OPTIMAS Research Center, RPTU University Kaiserslautern-Landau, 67663 Kaiserslautern, Germany\looseness=-1}%
\author{R. Ernstorfer}
\affiliation{Institut für Physik und Astronomie, Technische Universität Berlin, Straße des 17 Juni 135, 10623 Berlin, Germany\looseness=-1}%
\affiliation{Fritz Haber Institute of the Max Planck Society, Faradayweg 4--6, 14195 Berlin, Germany\looseness=-1}%
\author{T. Pincelli}
\email{pincelli@tu-berlin.de}
\affiliation{Institut für Physik und Astronomie, Technische Universität Berlin, Straße des 17 Juni 135, 10623 Berlin, Germany\looseness=-1}%
\affiliation{Fritz Haber Institute of the Max Planck Society, Faradayweg 4--6, 14195 Berlin, Germany\looseness=-1}%
\author{L. T. Lloyd}
\affiliation{Fritz Haber Institute of the Max Planck Society, Faradayweg 4--6, 14195 Berlin, Germany\looseness=-1}%

\maketitle

\section*{Additional Figures}

\renewcommand{\thefigure}{S1}
\begin{figure}[htb]
\centering
\includegraphics[width=0.95\linewidth]{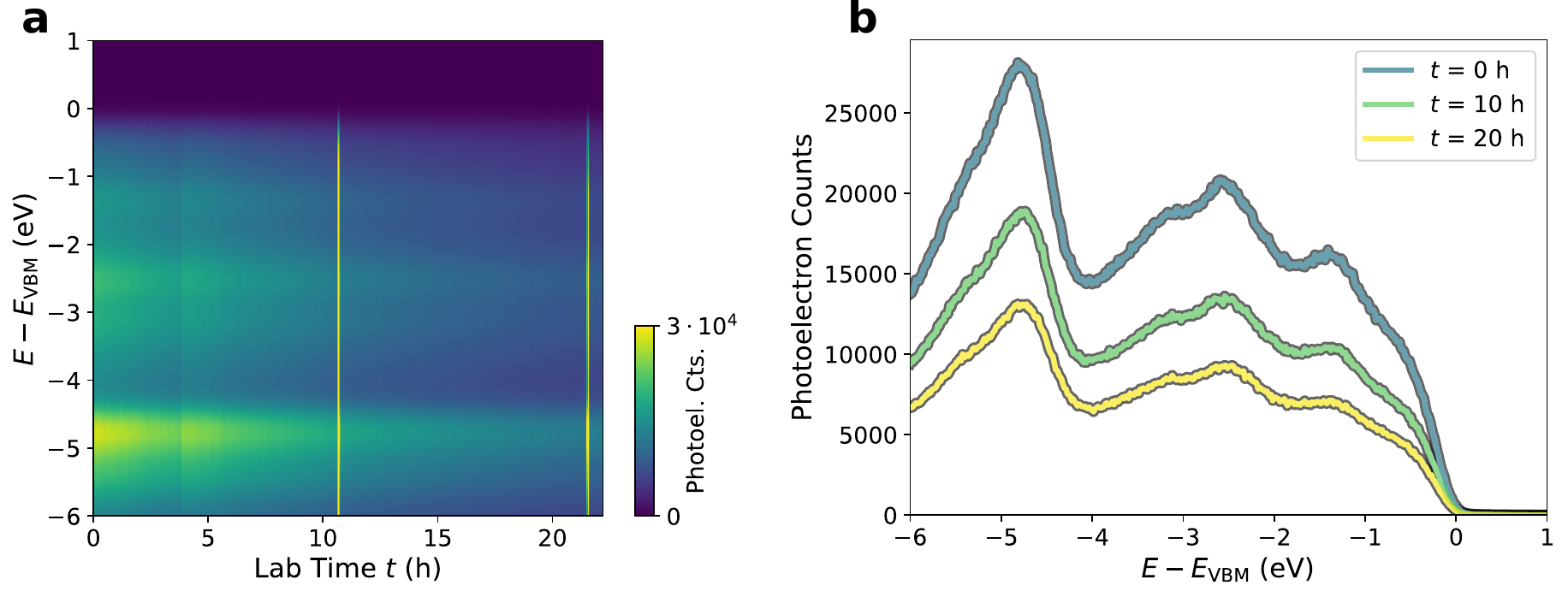}
\caption[captionsep=none]{\protect\label{fig:S1} \textbf{a}. \textit{k}-integrated photoelectron counts of NiI$_2$ over the course of a $\approx\SI{20}{\hour}$ measurement at \SI{300}{\K}. \textbf{b}. Individual EDCs of the photoemission distribution in \textbf{a} after \SI{0}{\hour}, \SI{10}{\hour}, and \SI{20}{\hour} of measurement.}
\end{figure}

In Fig.~\ref{fig:S1} we display the evolution of the sample conditions across an experimental run of $\approx20$ hours, as monitored from the \textit{k}-integrated spectrum. Fig.~\ref{fig:S1}a shows the energy distribution of the photoemitted intensity (represented by the color scale) as a function of time. The lineshape of the spectra doesn't change over the course of the whole experiment lasting $\sim$24 hours, as demonstrated by sampling in three positions (Fig.~\ref{fig:S1}b). Apart from a general decrease in photoemitted counts over time, we detect no appreciable change, demonstrating the robustness of our ARPES dataset over time. The gradual decrease in intensity has been accounted for in time-resolved measurements by continuously scanning the pump delay line in both directions, so to average changes in intensity over time across one cycle. 

\renewcommand{\thefigure}{S2}
\begin{figure}[htb]
\centering
\includegraphics[width=0.95\linewidth]{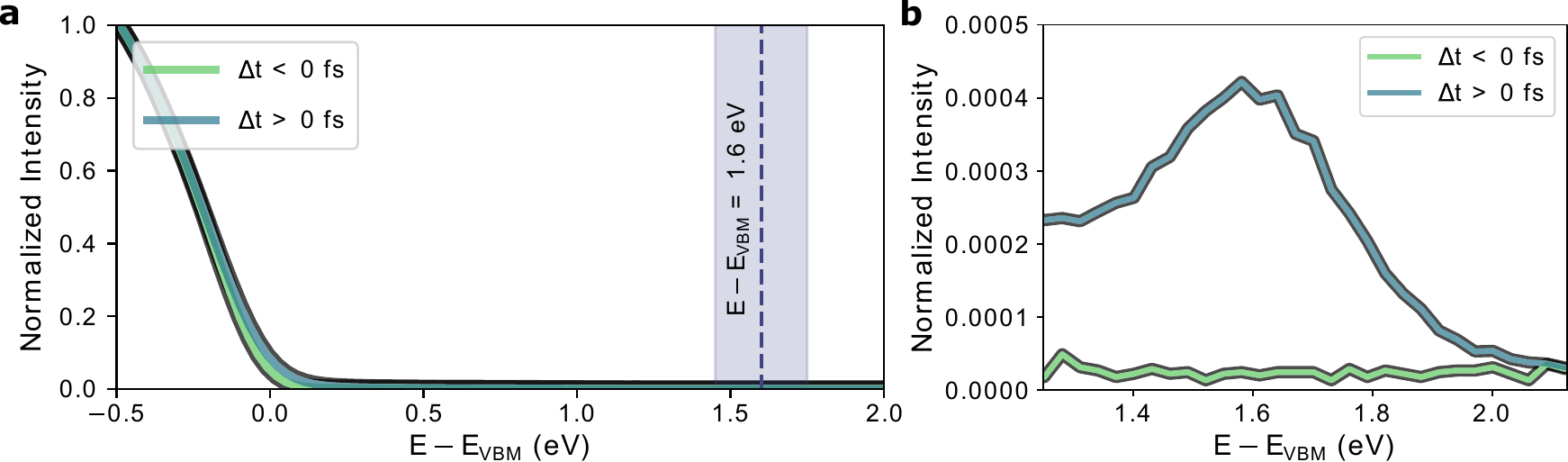}
\caption[captionsep=none]{\protect\label{fig:S2} \textbf{a}. EDCs before and after pump-probe time overlap, in the range including both occupied and excited states (a) and excited states only (b). The curve at negative times has been integrated from \SIrange{-250}{-150}{\fs}, the one at positive times from \SIrange{100}{400}{\fs}.}
\end{figure}

\hl{In Fig.~\ref{fig:S2} we show the comparison between EDCs at negative and positive time delays. The graphs display a different energy range, including both occupied and excited states (a) and excited states only (b), and different scaling for the vertical axis: there is a negligible pump-induced effect within the scale of occupied states, with only a minor shift (a), and no excited state signal at negative time delays (b). This confirms the absence of contributions to the photoemitted signal from the CW laser used to compensate sample charging.}

In Fig.~\ref{fig:S3} we plot the DFT calculations, overlaid upon the experimental ARPES (\textit{E}, \textit{k}) spectrum, performed with varying \textit{U} and \textit{J} values on a NiI$_2$ monolayer. We can note how increasing \textit{U} pushes the Ni 3\textit{d} manifold of weakly dispersive bands found between \SI{-3}{\eV} and \SI{-5}{\eV} towards higher binding energies. On the other hand, \textit{J} acts on all bands, and changes significantly the location of the I 5\textit{p} states near the high-symmetry point at the edges of the Brillouin zone (BZ). 

We strike a balance with a \textit{U} value that matches the texture of the Ni bands at the bottom of the spectrum, and with a \textit{J} value that reproduces the energy position and dispersion of the high-intensity hotspots at lower binding energies. The calculations discussed in the main text are based off these values.

\renewcommand{\thefigure}{S3}
\begin{figure}[htb]
\centering
\includegraphics[width=0.95\linewidth]{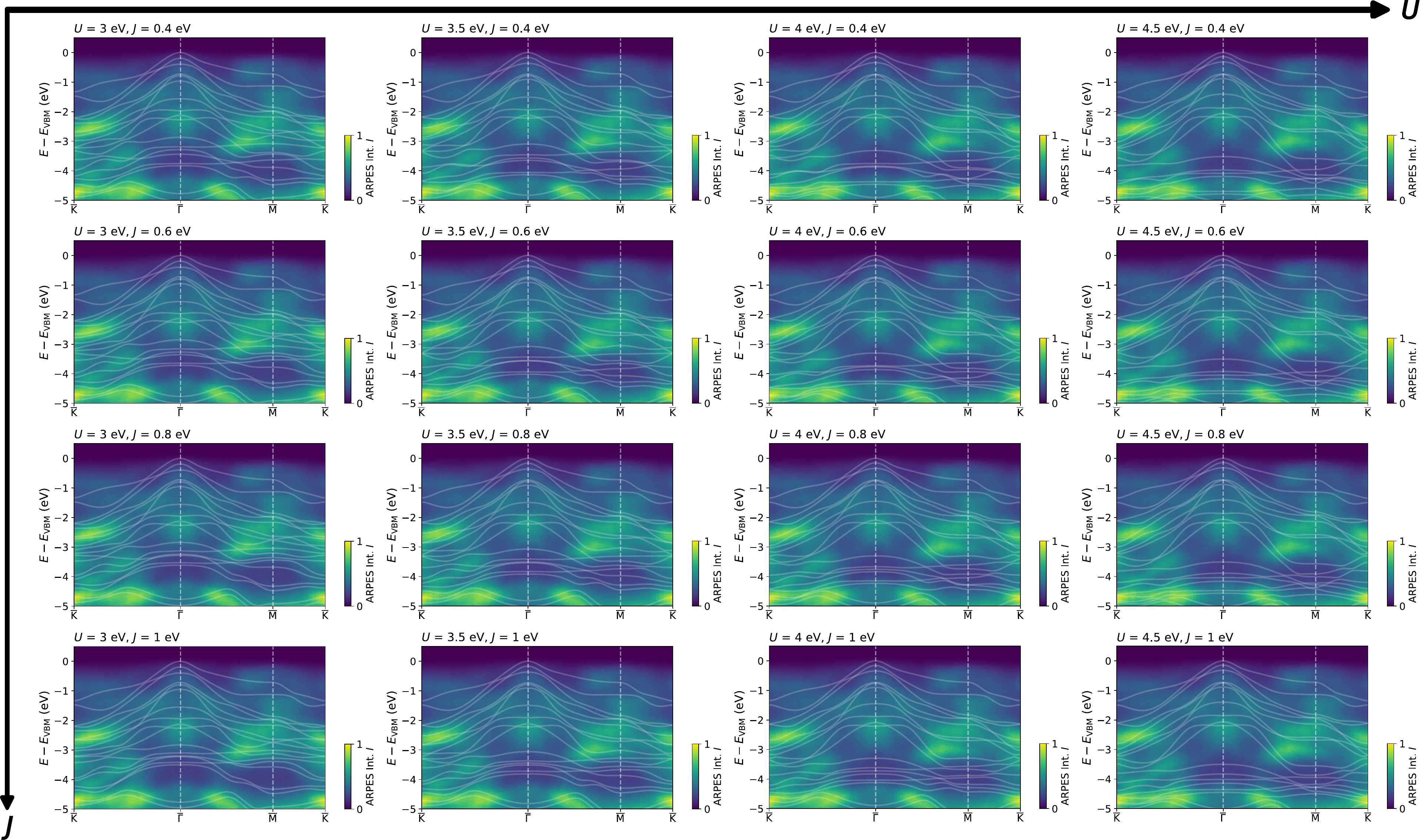}
\caption[captionsep=none]{\protect\label{fig:S3} Band structure of monolayer NiI\textsubscript{2}, computed for a set of \textit{U} and \textit{J} values on the Ni 3\textit{d} orbitals, overlaid on top of the experimental ARPES spectrum in Fig.~1d of the main text. The best match with experimental data was found for $U=\SI{3.5}{\eV}$ and $J=\SI{0.6}{\eV}$, and these are the values employed in the main paper.}
\end{figure}

\renewcommand{\thefigure}{S4}
\begin{figure}[htb]
\centering
\includegraphics[width=0.95\linewidth]{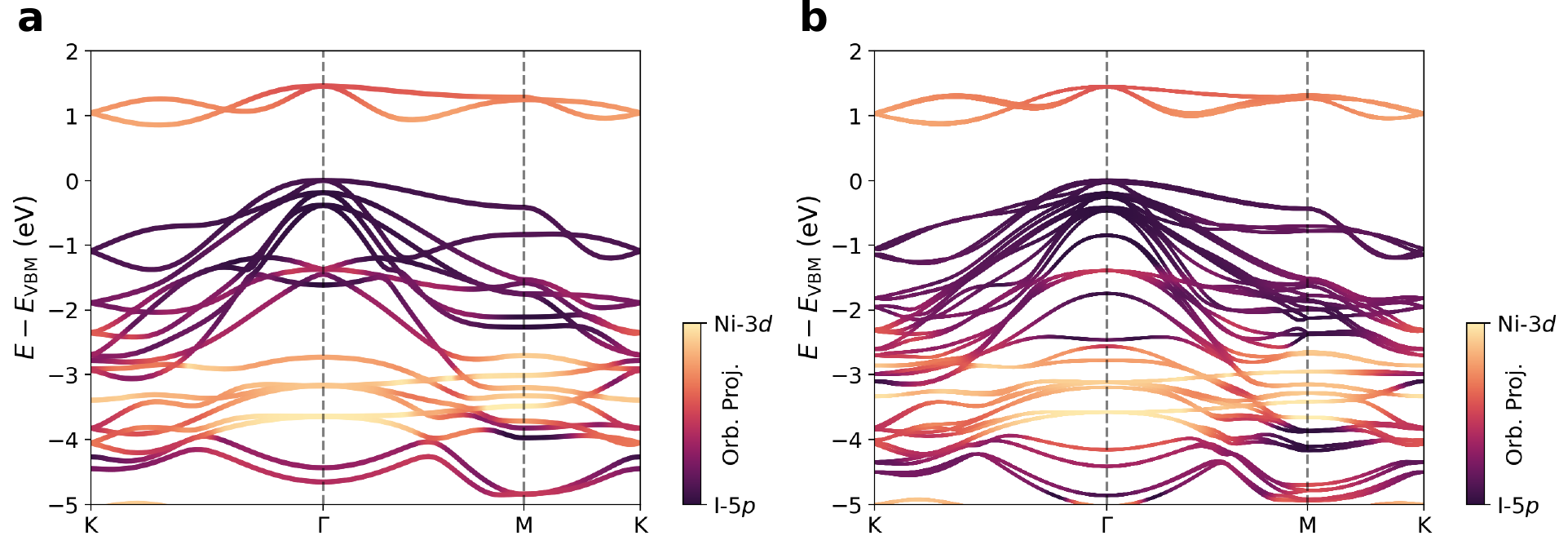}
\caption[captionsep=none]{\protect\label{fig:S4} Band structure of \textbf{a}. monolayer and \textbf{b}. bulk NiI\textsubscript{2}, calculated with \hl{$U=\SI{3.5}{\eV}$} and $J=\SI{0.6}{\eV}$.}
\end{figure}

To justify the assumption of a good match between the experiment and our monolayer calculations, in Fig.~\ref{fig:S4} we display a comparison between collinear calculations on monolayer and bulk NiI$_2$. We confirm that the general character of the bands in terms of position and dispersion is very similar between the two cases, with a number of additional states being present for the bulk calculation, as expected. The most striking difference, as mentioned in the main text, is found between \SIrange{-2}{-3}{\eV} at the BZ center, where extra hole-like bands appear. Other manifolds display minor dissimilarities in the dispersion, but follow broadly the same trends, vindicating our previous hypothesis.

\renewcommand{\thefigure}{S5}
\begin{figure}[htb]
\centering
\includegraphics[width=0.95\linewidth]{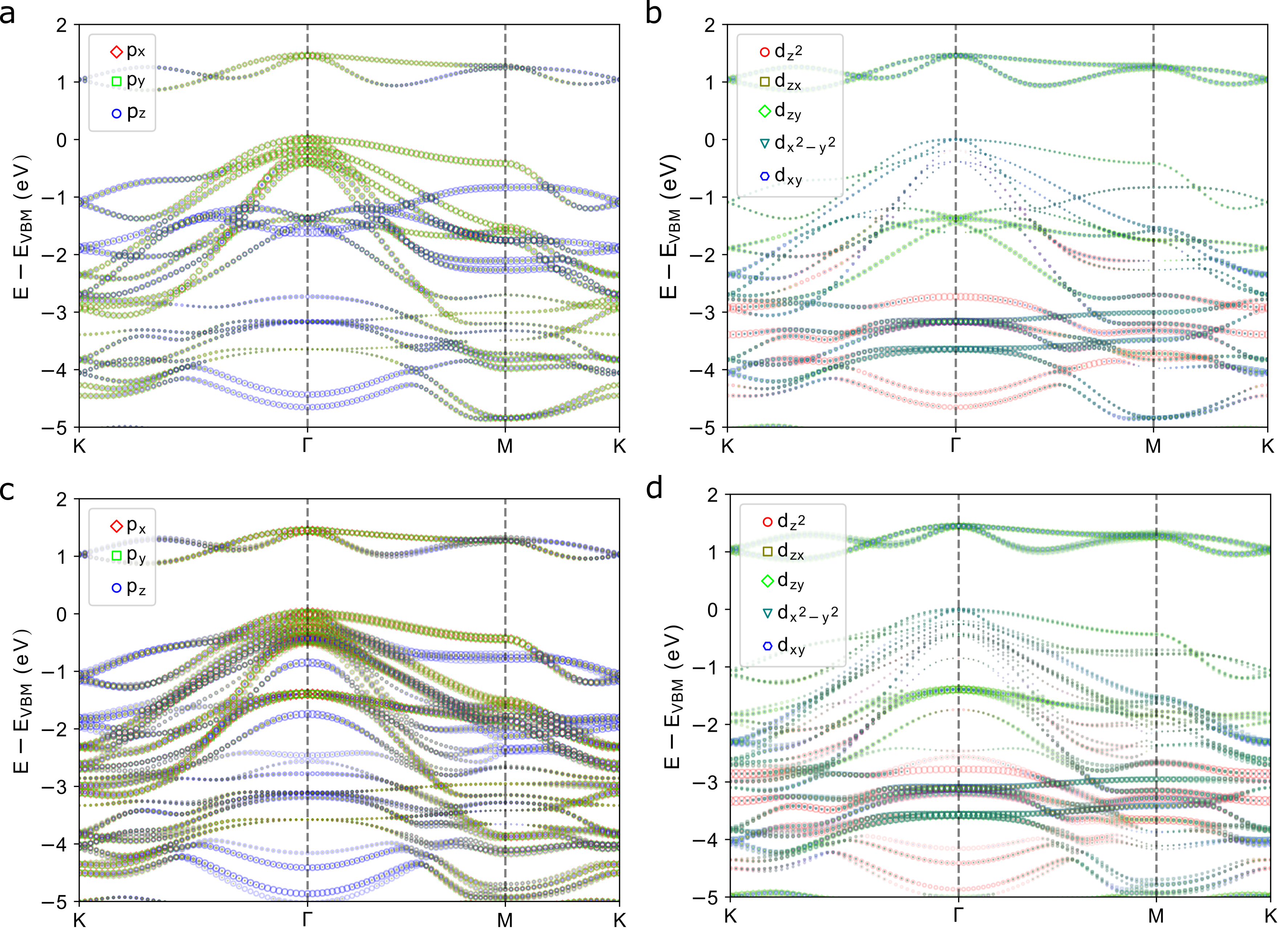}
\caption[captionsep=none]{\protect\label{fig:S5} Orbital-projected DFT calculations of ML (\textbf{a,b}) and bulk (\textbf{c,d}) NiI$_2$. Left and right panels display the \textit{p}-orbital and \textit{d}-orbital projections, respectively.}
\end{figure}

\hl{Additionally, we computed the projection of the calculations in Fig.~\ref{fig:S4} onto atomic orbitals. Results are shown in Fig.~\ref{fig:S5}. We compare these calculations with experimental spectra shown in Fig.~\ref{fig:S6}, where the differences between ARPES spectra acquired with \textit{s}-polarized and \textit{p}-polarized light are highlighted. In particular, as discussed in the main text, the valence band maximum at $\bar{\Gamma}$, revealed to be mostly visible with \textit{s}-polarized photons, originates from orbitals with $p_x/p_y$ symmetry; conversely, states at $\bar{\Gamma}$ but at higher binding energy display strong out-of-plane character, whereas the dispersions at the outskirts of the Brillouin zone show a more mixed origin.} 

\renewcommand{\thefigure}{S6}
\begin{figure}[htb]
\centering
\includegraphics[width=0.95\linewidth]{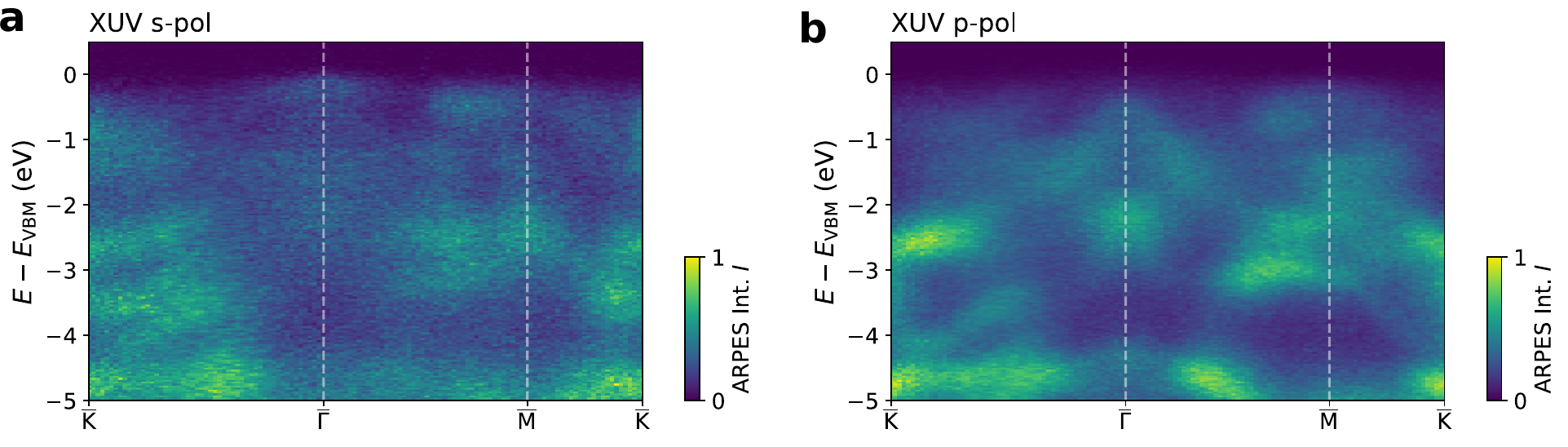}
\caption[captionsep=none]{\protect\label{fig:S6} ARPES (\textit{E}, \textit{k}) spectra of NiI\textsubscript{2} at \SI{300}{\K}, acquired with XUV photons at \SI{21.7}{\eV} at \textbf{a}. \textit{s}-polarization, and \textbf{b}. \textit{p}-polarization configurations.}
\end{figure}

\hl{In Fig.~\ref{fig:S7} we plot ARPES measurements of bulk NiI$_2$ at \SI{30}{\K}, to be compared with room-temperature measurements as presented in the main text. The measurement in Fig.~\ref{fig:S7} has been performed on the same sample and in the same azimuthal geometry as the room-temperature measurements.}

\renewcommand{\thefigure}{S7}
\begin{figure}[htb]
\centering
\includegraphics[width=0.8\linewidth]{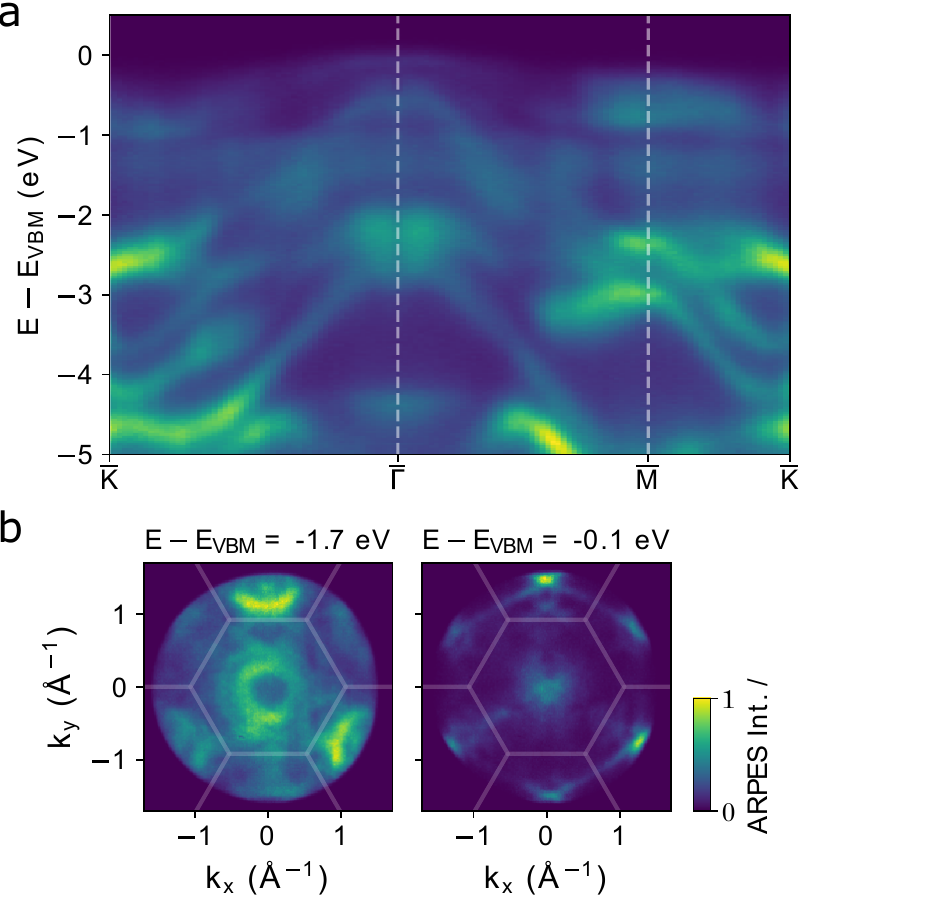}
\caption[captionsep=none]{\protect\label{fig:S7} ARPES static (\textit{E}, \textit{k}) spectra (\textbf{a}) and momentum maps (\textbf{b}) of bulk NiI\textsubscript{2} at \SI{30}{\K}, acquired with XUV photons at \SI{21.7}{\eV} and variable polarization, as in Fig.~1d of the main text. Momentum maps are acquired at \SI{-1.7}{\eV} and \SI{-0.1}{\eV} respectively, with an integration window of \SI{0.1}{\eV}.}
\end{figure}

\hl{Figs.~~\ref{fig:S8} and ~\ref{fig:S9} provide additional details on the excited states, respectively about the momentum distribution via MDCs and about the dependence of the excitation on the pump polarization. }

\renewcommand{\thefigure}{S8}
\begin{figure}[htb]
\centering
\includegraphics[width=0.55\linewidth]{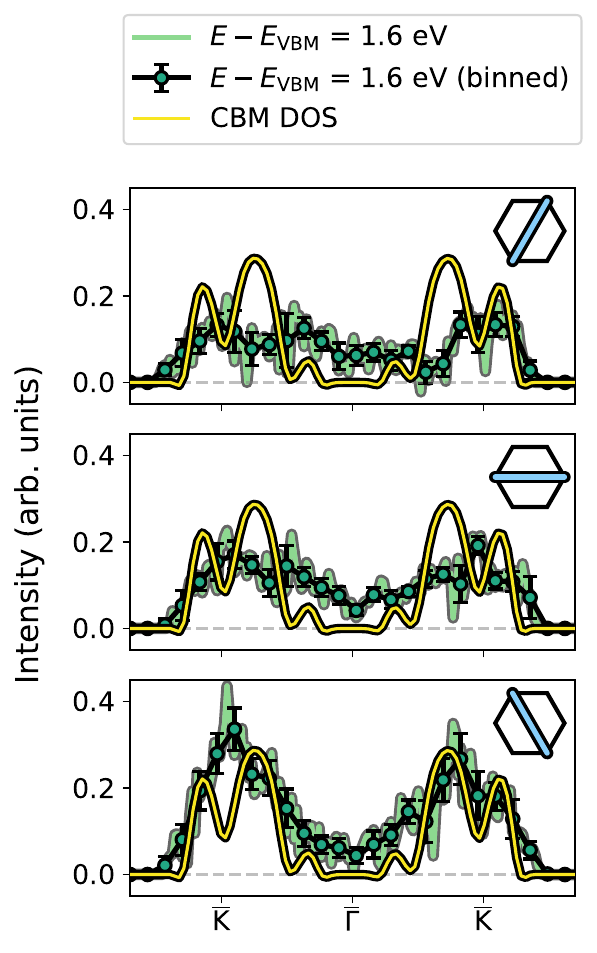}
\caption[captionsep=none]{\protect\label{fig:S8} MDCs extracted along the $\bar{K}-\bar{\Gamma}-\bar{K}$ directions at \SI{1.6}{\eV} (excited state), plotted as raw data and binned with a factor 4, together with the corresponding DOS extracted from DFT calculations. The energy integration window is \SI{0.3}{\eV}. The inset displays which $\bar{K}-\bar{\Gamma}-\bar{K}$ direction is represented.}
\end{figure}

\renewcommand{\thefigure}{S9}
\begin{figure}[htb]
\centering
\includegraphics[width=0.95\linewidth]{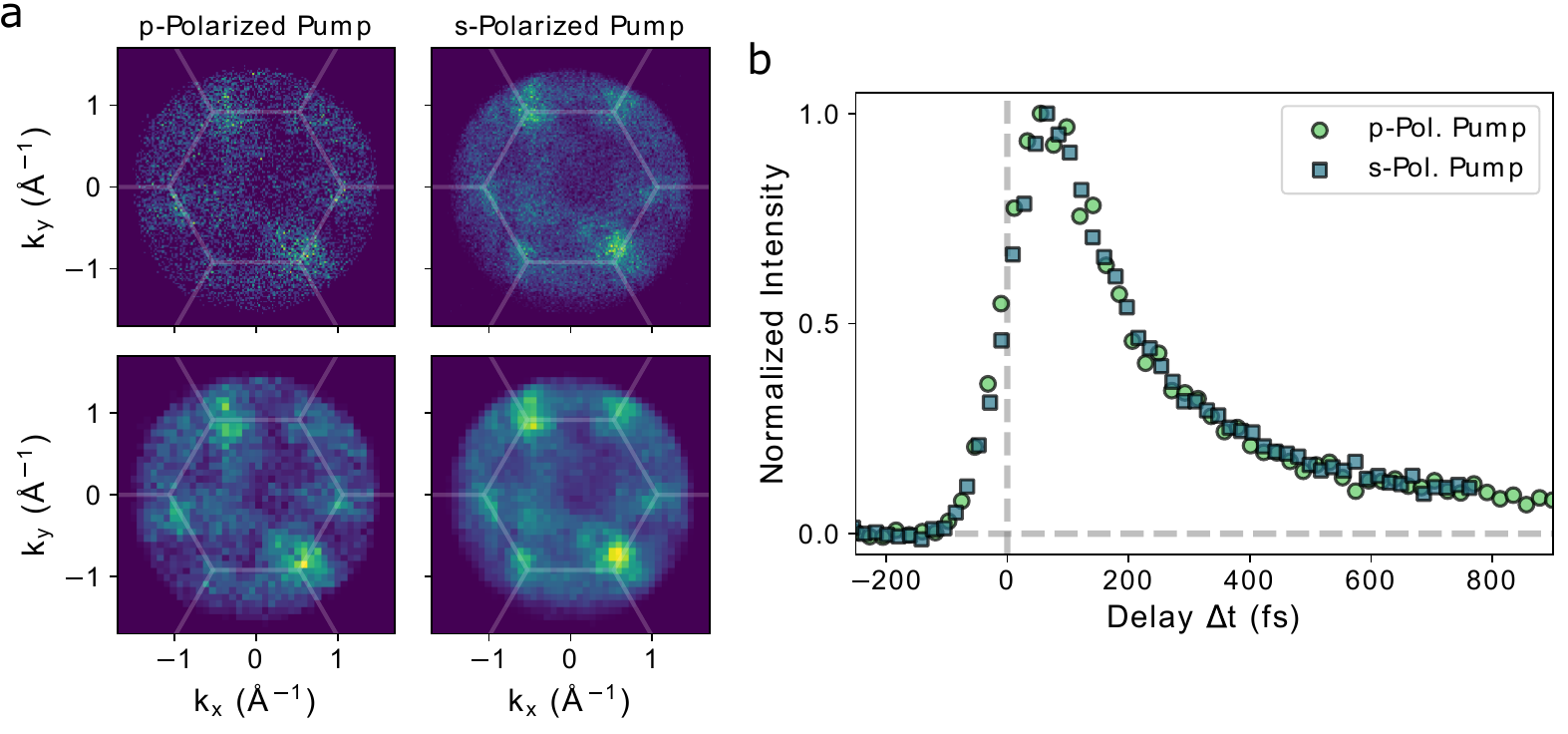}
\caption[captionsep=none]{\protect\label{fig:S9} \textbf{a}. ARPES momentum maps of the excited state at \SI{1.6}{\eV}, measured with \textit{p}- and \textit{s}-polarized pump. The top row shows raw data, the bottom row displays binned data (factor 4). Maps are integrated in a \SI{0.3}{\eV} energy window. \textbf{b}. Delay curves of the excited state signal integrated in the same energy range, for both pump polarizations, normalized to the intensity maximum.}
\end{figure}

%\bibliography{bibliography}